\documentclass[pra,aps,twocolumn]{revtex4}
\usepackage{amsmath,amsfonts,latexsym,graphicx,amssymb}
\def\dST{\displaystyle}
\newtheorem{example}{Example}
\newtheorem{theorem}{Theorem}

\def\1{{\mathchoice{\rm 1\mskip-4mu l}{\rm 1\mskip-4mu l}%
{\rm 1\mskip-4.5mu l}{\rm 1\mskip-5mu l}}}
\def\<{\langle}
\def\>{\rangle}

\def\dST{\displaystyle}
\def\ket#1{|\kern0.7pt #1\>}
\def\bra#1{\langle #1\kern0.7pt|}
\def\scalar#1#2{\langle#1\kern0.7pt|\kern0.7pt #2\rangle}
\def\Tr{{\rm Tr}\kern1pt}
\begin{document}
\title{Estimating Concurrence via Entanglement Witnesses}
\author{Jacek Jurkowski and Dariusz Chru\'sci\'nski}
\affiliation{Institute of Physics, Nicolaus Copernicus University\\
 ul. Grudzi\c{a}dzka 5/7, 87--100 Toru\'n, Poland}

\begin{abstract}
We show that each entanglement witness detecting given bipartite
entangled state provides an estimation of its concurrence. We
illustrate our result with several well known examples of
entanglement witnesses and compare the corresponding estimation of
concurrence with other estimations provided by the trace norm of
partial transposition and realignment.
\end{abstract}


\maketitle
%\numberwithin{equation}{section}

\section{Introduction}

 The interest on  quantum entanglement has
dramatically increased during the last two decades due to the
emerging field of quantum information theory \cite{QIT,Hor09}. It
turns out that quantum entanglement may be used as basic resources
in quantum information processing and communication. The prominent
examples are quantum cryptography, quantum teleportation, quantum
error correction codes and quantum computation.
 Hence it is of basic
importance from both experimental and theoretical point of view to
provide methods of detecting and quantifying entanglement
\cite{Toth,Guh07,Eis}. There are no universal criteria to detect quantum
entanglement and there are few measures of entanglement (based on
the notion of entropy \cite{Ved98}, entanglement of formation
\cite{Ben96}, concurrence \cite{Woo98}, robustness
\cite{Vid,Steiner}, geometrical measures \cite{Bar01,Wei03} and
others).  They can be calculated for pure quantum states or  for the
very limited class of  mixed states possessing  some symmetry
properties \cite{isotropic,KaiRMP,Ter,Loh}. Therefore, the great
effort is directed to obtain methods of estimation of particular
entanglement measures and to find relations between them
\cite{Datta,Guh08,AugLew09,Ma,deVicente,Min04}.

On the other hand, what is measured in an experiment it is an
expectation value of some observables, hence the estimations based
on  such quantities are most welcome \cite{Min07,Min07a,Min07b}. For
example, it was recently shown \cite{Min07b} that the concurrence
for pure bipartite state $\ket{\psi}$ living in a Hilbert space
${\cal H}$ can be obtained as follows
\[
C(\ket{\psi})\;=\;2\sqrt{\bra{\psi}\otimes\bra{\psi}\hat{\Pi}\ket{\psi}\otimes\ket{\psi}}\,,
\]
where $\hat{\Pi}=\hat{\Pi}_+\otimes\hat{\Pi}_+$ and $\hat{\Pi}_+$ is the projector onto
the symmetric subspace of ${\cal H}$, i.e., concurrence is expressible
by the mean value of the observable $\hat\Pi$ acting on the two-copy space ${\cal H}
\otimes{\cal H}$.

It turns out that useful candidates for this  purpose are
entanglement witnesses \cite{Min07a} (see \cite{Chr} for the recent review on entanglement witnesses).
We shown that each
entanglement witness detecting given bipartite entangled state in
$\mathcal{H}_A \otimes \mathcal{H}_B$ provides an estimation of its
concurrence. Hence, EWs define a universal tool not only for
detecting quantum entanglement but also for estimating its measure.
We compare estimation based on entanglement witnesses with other
ones provided by the trace norm of partial transposition and
realignment.

The paper is organized as follows: in the next Section we provide
basic introduction to concurrence and its estimations. Section
\ref{MAIN} presents our main result which is illustrated by the
family of examples in Section \ref{EX}.   Final conclusions are
collected in the last section.

\section{Concurrence -- preliminaries}

Let us recall that the  concurrence  for a pure bipartite state
$\ket{\psi}\in {\cal H}_{A}\otimes{\cal H}_B$ is defined as follows
\begin{equation}\label{}
C(\ket{\psi})\;=\;\sqrt{2(1-\Tr\rho_A^2)}\,,
\end{equation}
where $\rho_A=\Tr_B(\ket{\psi}\bra{\psi})$ is the reduced density
matrix. In the following, we will use a Schmidt decomposition of the
pure state
            \begin{equation}\label{Schm-decomp}
            \ket{\psi}\;=\;\sum_{i=1}^m\sqrt{\mu_i}\ket{a_i}\otimes\ket{b_i}
            ,
            \end{equation}
where $m = \min\{ {\rm dim}\mathcal{H}_A,{\rm dim}\mathcal{H}_B\}$,
and
 $\{\ket{a_i}\}$, $\{\ket{b_i}\}$ are orthonormal bases in ${\cal H}_A$
and ${\cal H}_B$, respectively. The  Schmidt coefficients $\mu_i\geq
0$ and satisfy the following normalization condition
\begin{equation}\label{}
\sum_{i=1}^m\mu_i\;=\;1\,.
\end{equation}
It is easy to check that concurrence $C(\ket{\psi}))$ is uniquely
defined in terms of the Schmidt coefficients $\mu_i$. One has
\begin{equation}\label{}
 C(\ket{\psi})\;=\;\sqrt{2\sum_{k,l\neq k}\mu_k\mu_l}\,.
\end{equation}
For a mixed state $\rho=\sum_ip_i\ket{\psi_i}\bra{\psi_i}$ the
concurrence is defined via a convex roof construction
\begin{equation}\label{mixed-state-conc}
    C(\rho)\;=\;\min_{\{p_i,\ket{\psi_i}\}}\sum_ip_iC(\ket{\psi_i})\,.
\end{equation}
It is well known \cite{Woo98} that for a two-qubit case one finds
the following formula for the concurrence of the arbitrary mixed
state
\begin{equation}\label{}
C(\rho)\;=\;\max\{0,\lambda_1-\lambda_2-\lambda_3-\lambda_4\},
\end{equation}
where $\lambda_1\geq\lambda_2\geq\lambda_3\geq\lambda_4$ are
singular values of a matrix $T_{kl}=\bra{v_k}\sigma_y
\otimes\sigma_y\ket{v_l^*}$ with $\ket{v_k}$ denoting eigenvectors
of $\rho$ and $\sigma_y$ stands for the Pauli matrix. In general,
however, one has only the following estimation \cite{Ou}
\begin{equation}\label{}
C^2(\rho)\;\geq\;\sum_{k=1}^{D}\sum_{l=1}^{D}
\Big(\max\{0,\lambda^{(1)}_{kl}-\lambda^{(2)}_{kl}-
\lambda^{(3)}_{kl}-\lambda^{(4)}_{kl}\}\Big)^2\ ,
\end{equation}
where now $\lambda^{(i)}_{kl}$ are singular values of
$(T^{k,l})_{\alpha,\beta}=\bra{v_\alpha}L_k\otimes L_l\ket{v_\beta^*}$, $D=m(m-1)/2$ and
$L_k$ are generators of the SO$(m)$ group.
It is also possible to carry out the optimalisation procedure involved in (\ref{mixed-state-conc})
for particular families of states possessing some symmetry properties (Werner states,
isotropic states) \cite{isotropic,KaiRMP,Ter,Loh}.

       Let us recall two basic results which enable estimation of concurrence for an
        arbitrary entangled mixed state $\rho$.
        % when either
        %the trace norm of the partial transposed state or the trace norm of a realigned matrix corresponding to $\rho$ is known.
        \par\smallskip\noindent
\begin{theorem}[Chen, Albeverio, Fei \cite{Chen}]
 The following estimation is valid:
        \begin{equation}\label{C-thm1}
        C(\rho) \geq  \sqrt{\frac{2}{m(m-1)}}\Big(\max\{||\rho^{T_A}||_1,||{\cal
        R}(\rho)||_1\}-1\Big) \ ,
        \end{equation}
        where $|| X ||_1$ denotes the trace norm of $X$.
\end{theorem}
        \par\smallskip\noindent
As usual $\rho^{T_A}$ denotes a partial transposition of $\rho$ and
${\cal R}(\rho)$ stands for the  realigned matrix \cite{R1,R2}. For some generalizations see \cite{Zhang}.
        Note that although for a PPT state $||\rho^{T_A}||_1=1$, the norm of a realigned matrix
        $||{\cal R}(\rho)||_1$ can still be greater than 1 resulting in a nontrivial estimation.

        Let us recall that a hermitian operator $W$ is called an entanglement witness for a state $\rho$, if
        ${\rm Tr}(\rho W)<0$ while ${\rm Tr}(\sigma W)\geq 0$ for all separable states $\sigma$.
        There are many examples \cite{Brandao} of entanglement measures $M(\rho)$ (concurrence, negativity, robustness, etc.)
        which can be related to the expectation value of some entanglement witness
\begin{equation}\label{}
M(\rho)\;=\;\max\Big\{0,-\inf_{W\in{\cal M}}\Tr(\rho W)\Big\}\,,
\end{equation}
where the set ${\cal M}$ depends on the measure $M$ in question. It
is therefore clear that if $W$ is an entanglement witness for
$\rho$, i.e.~$\Tr(\rho W)<0$, and $W\in {\cal M}$, then one finds
the following estimation
\[ M(\rho)\;\geq\;|\Tr(\rho W)|
\]
for the measure of entanglement of $\rho$. In the case of
concurrence one has the following theorem

\begin{theorem}[Breuer \cite{Breuer}]  \label{T-B}
         Let $W$ be an entanglement witness such that
        \begin{equation}\label{cond}
         -\bra{\psi}W\ket{\psi}\;\leq\;\sum_{i,j\neq i}\sqrt{\mu_i\mu_j}
        \end{equation}
        for every pure state (\ref{Schm-decomp}).
        Then for an arbitrary mixed state $\rho$ detected by the
        witness $W$
        \begin{equation}\label{witness-condition}
        C(\rho) \;\geq\; \sqrt{\frac{2}{m(m-1)}}\ |\Tr(\rho W)|\,.
        \end{equation}
\end{theorem}

\section{Main result} \label{MAIN}

Theorem \ref{T-B} distinguishes a class of witnesses satisfying
condition (\ref{cond}). Suppose now that  $W$ does not satisfy this
condition. Clearly, for any $\alpha > 0$ the rescaled operator
$\alpha^{-1} W$ still defines an EW. Does $\alpha^{-1} W$ satisfy
(\ref{cond})? To answer this question let us observe that
    for $\ket{\psi}\;=\;\sum_{i=1}^m\sqrt{\mu_i}\ket{a_i,b_i}$ the expectation value of $W$
    reads as follows
\begin{equation}\label{}
\bra{\psi}W\ket{\psi}\;=\;\sum_{k,l}\sqrt{\mu_k\mu_l}A_{kl}^{(W)}(\psi)\,,
\end{equation}
where the $\psi$-dependent matrix $A_{kl}^{(W)}$ is defined by
    \begin{eqnarray}
    A_{kl}^{(W)}(\psi) =  {\rm Re}\,\bra{a_k,b_k}W\ket{a_l,b_l} \ .
    \end{eqnarray}
Note, that
\begin{equation}\label{kk}
 A_{kk}^{(W)}(\psi) \geq 0 \ ,
\end{equation}
by the very definition of entanglement witness. It is clear that $
A_{kl}^{(W)}(\psi)$ encodes the entire information about $W$.
Moreover, the condition (\ref{cond}) is equivalent to
\begin{equation}\label{new}
    \sum_{k,l}\sqrt{\mu_k\mu_l}(A_{kl}^{(W)}+1)\;\geq\; 1\,.
\end{equation}
Let us observe that the space of normalized vectors defines a
compact set and hence one may define a positive number $\lambda$ by
the following  procedure
\begin{equation}\label{lambda}
    - \lambda := \min_\psi \min_{k\neq l} A_{kl}^{(W)}(\psi) \ .
\end{equation}
 Now,
comes the main result
\begin{theorem} \label{T-M}
For any $\alpha \geq \lambda$ the rescaled entanglement witness
$\alpha^{-1} W$  does satisfy (\ref{cond}).
\end{theorem}

\noindent The proof is almost trivial. One has
   \begin{eqnarray*}
   \sum_{k,l}\sqrt{\mu_k\mu_l}(A_{kl}^{(W)}(\psi) +1) &=& 1+\sum_k\mu_kA_{kk}^{(W)}(\psi) \\
   &+&   \sum_{k,l\neq k}\sqrt{\mu_k\mu_l}(A_{kl}^{(W)}(\psi) +1) \\
   &\geq&1+\sum_{k,l\neq k}\sqrt{\mu_k\mu_l}(A_{kl}^{(W)}(\psi) +1)\ ,
   \end{eqnarray*}
where we have used (\ref{kk}). Hence, if
\begin{equation}\label{=}
 A_{kl}^{(W)}(\psi) \geq -1\ ,
\end{equation}
for every normalized $\psi$, then $W$ does satisfy (\ref{cond}).
Suppose now that the above condition is not satisfied. It is
therefore clear that for the rescaled witness $W_\alpha:=
\alpha^{-1} W$ with $\alpha \geq \lambda$,  one has
\begin{equation}\label{optimal}
 A_{kl}^{(W_\alpha)}(\psi) \geq -1\ ,
\end{equation}
which proves our theorem. It should be stressed that the best estimation is provided by the witness corresponding to $\alpha=\lambda$.

\section{Examples}  \label{EX}

\begin{example}
{\em Let $\mathcal{H}_A = \mathcal{H}_B = \mathbb{C}^m$ and consider
the flip operator
   \[ F\;=\;\sum_{i,j=1}^m \ket{i}\bra{j}\otimes\ket{j}\bra{i} \]
where $\{\ket{i}\}$ is the computional basis in $\mathbb{C}^m$.
Simple calculation gives
   \[ A_{kl}^{(F)}\;=\;\bra{a_k,b_k}F\ket{a_l,b_l}\;=\;\scalar{a_k}{b_l}
   \scalar{b_k}{a_l}\,. \]
   Now, evidently $A_{kk}^{(F)}=|\scalar{a_k}{b_k}|^2\geq0$ and for $k\neq l$
   \[ A_{kl}^{(F)}\;=\;{\rm Re}\,(\scalar{a_k}{b_l}
   \scalar{b_k}{a_l})\;\geq\;-1
   \]
   according to orthonormality of both basis. }
\end{example}
\begin{example}\rm
Let us consider isotropic states in $\mathbb{C}^m\times\mathbb{C}^m$
\begin{equation}\label{isotropic}
\rho_f=\frac{1-f}{m^2-1}(\1-P_m^+)+f P_m^+\,,
\end{equation}
where $P^+_m$ denotes the maximally entangled state and $f$ is the
fidelity defined by $f=\<\psi_m^+|\rho_f|\psi_m^+\>$. Moreover, one
introduces a family of $k$-EWs \cite{TerHor}
\begin{equation}
W^{\rm iso}_k=\frac{k}{m}\1-P_m^+\,,\qquad k=1,\ldots,m-1\,,
\end{equation}
satisfying
\[ \Tr[W^{\rm iso}_k\rho_f]=\frac{k}{m}-f\,,
\]
that is, $W^{\rm iso}_k$ detects isotropic state with fidelity $f
> k/m$. Such state has Schmidt number strictly greater than $k$. Since $P_m^+=\frac1m
F^{T_A}$  the previous example implies for $i\neq j$
\[
A_{ij}^{(W^{\rm iso}_k)}\geq -\frac{1}{m}\,,
\]
which shows that $\lambda=1/m$. As a consequence, the optimal $W^{\rm
iso}_k$, in the sense of (\ref{optimal}), is $\widetilde{W}^{\rm
iso}_k=mW^{\rm iso}_k$. Now,
\begin{equation}
\Tr(\rho_f\widetilde{W}^{\rm iso}_k)=m\Tr(\rho_fW^{\rm iso}_k)=
m\Big(\frac{k}{m}-f\Big)
\end{equation}
and the estimation (\ref{witness-condition}) takes the form
\[ \sqrt{\frac{2m}{m-1}}\Big(f-\frac{k}{m}\Big)\leq C(\rho_f)\,.
\]
Note that although for $k\neq 1$, $W^{\rm iso}_k$ provides only the bound for concurrence,
when $k=1$, we obtain an {\it exact\/} result \cite{isotropic}.
\end{example}

Let us note that  a much more general (but also numerically more
involved) method of estimating various entanglement measures was
proposed in \cite{Guh08}. The method uses a concept of  an
entanglement witness and on the other hand provides a numerical
procedure to calculate the Legendre transform of the measure in
question. The method provided in this paper is much more restricted.
However, being simpler it provides estimation of concurrence which
can be very often computed analytically. The above examples show
that it can leads not only to upper bounds for concurrence but also
to exact results.

\begin{example}
{\em In \cite{Jur09} we have investigated an $\varepsilon$-family
($\varepsilon > 0$) of states in $\mathbb{C}^3 \otimes \mathbb{C}^3$
   \begin{equation}\label{rhoE}
   \rho(\varepsilon)\;=\;\ N_\varepsilon\Big( P^+_3 + \frac 13 \sum_{i\neq j=1}^3 d_{ij}\ket{ij}\bra{ij} \Big) \,,
   \end{equation}
where $P^+_3$ denotes a maximally entangled state,
\[ d_{i,i+1}=\varepsilon\,,\qquad d_{i,i+2}=\frac1{\varepsilon}
\ , \ \ \ (mod \ 3) \] and the normalization factor
\[  N_\varepsilon = \frac{1}{1 + \varepsilon + \varepsilon^{-1}} \ .
\]
It turns out that $\rho(\varepsilon)$ is entangled if and only if
$\varepsilon \neq 1$. Moreover, its entanglement is detected  by the
entanglement witness
\begin{equation}\label{choi1}
W_1\;=\;
\left(\begin{array}{ccc|ccc|ccc}
             1 & \cdot & \cdot & \cdot & -1 & \cdot & \cdot & \cdot & -1 \\
             \cdot & 1 & \cdot & \cdot & \cdot & \cdot & \cdot & \cdot & \cdot \\
             \cdot & \cdot & \cdot & \cdot & \cdot & \cdot & \cdot & \cdot & \cdot \\
\hline
             \cdot & \cdot & \cdot & \cdot& \cdot & \cdot & \cdot & \cdot & \cdot \\
             -1 & \cdot & \cdot & \cdot & 1 & \cdot & \cdot & \cdot & -1 \\
             \cdot & \cdot & \cdot & \cdot & \cdot & 1 & \cdot & \cdot & \cdot \\
\hline
             \cdot & \cdot & \cdot & \cdot & \cdot & \cdot & 1 & \cdot & \cdot \\
             \cdot & \cdot & \cdot & \cdot & \cdot & \cdot & \cdot & \cdot & \cdot \\
             -1 & \cdot & \cdot & \cdot & -1 & \cdot & \cdot & \cdot & 1
            \end{array}\right),
\end{equation}
for $\varepsilon<1$ and
\begin{equation}\label{choi2}
W_2\;=\;
\left(\begin{array}{ccc|ccc|ccc}
             1 & \cdot & \cdot & \cdot & -1 & \cdot & \cdot & \cdot & -1 \\
             \cdot & \cdot & \cdot & \cdot & \cdot & \cdot & \cdot & \cdot & \cdot \\
             \cdot & \cdot & 1 & \cdot & \cdot & \cdot & \cdot & \cdot & \cdot \\
\hline
             \cdot & \cdot & \cdot & 1& \cdot & \cdot & \cdot & \cdot & \cdot \\
             -1 & \cdot & \cdot & \cdot & 1 & \cdot & \cdot & \cdot & -1 \\
             \cdot & \cdot & \cdot & \cdot & \cdot & \cdot & \cdot & \cdot & \cdot \\
\hline
             \cdot & \cdot & \cdot & \cdot & \cdot & \cdot & \cdot & \cdot & \cdot \\
             \cdot & \cdot & \cdot & \cdot & \cdot & \cdot & \cdot & 1 & \cdot \\
             -1 & \cdot & \cdot & \cdot & -1 & \cdot & \cdot & \cdot & 1
            \end{array}\right)
\end{equation}
for $\varepsilon>1$. To make pictures more transparent we replaced
all zeros by dots. Interestingly, $W_1$ corresponds to the
celebrated Choi positive indecomposable map and $W_2$ to its dual.
Numerical calculations show that indeed $A_{kl}^{(W_i)}(\psi)\geq-1$
for $i=1,2$.
   %\rho(\varepsilon)\;=\;\left(\begin{array}{ccc|ccc|ccc}
   %1 &\cdot&\cdot&\cdot&1&\cdot&\cdot&\cdot&1 \\
   %\cdot&\varepsilon&\cdot&\cdot&\cdot&\cdot&\cdot&\cdot&\cdot \\
   %\cdot&\cdot&1/\varepsilon&\cdot&\cdot&\cdot&\cdot&\cdot&\cdot \\ \hline
   %\cdot&\cdot&\cdot&1/\varepsilon&\cdot&\cdot&\cdot&\cdot&\cdot \\
   %1 &\cdot&\cdot&\cdot&1&\cdot&\cdot&\cdot&1 \\
   %\cdot&\cdot&\cdot&\cdot&\cdot&\varepsilon&\cdot&\cdot&\cdot \\ \hline
   %\cdot&\cdot&\cdot&\cdot&\cdot&\cdot&\varepsilon&\cdot&\cdot \\
   %\cdot& \cdot&\cdot&\cdot&\cdot&\cdot&\cdot&1/\varepsilon&\cdot \\
   %1 &\cdot&\cdot&\cdot&1&\cdot&\cdot&\cdot&1
   %\end{array}\right)
Hence one obtains  the following estimation for concurrence based on
the above EWs
   \begin{equation}\label{exmp2}
     C(\rho(\varepsilon))\;\geq\;-\frac{1}{\sqrt{3}}\left\{\begin{array}{cl}
   \dST\frac{\varepsilon(\varepsilon-1)}{1+\varepsilon+\varepsilon^2} &\quad 0<\varepsilon<1 \\[2ex]
   \dST\frac{1-\varepsilon}{1+\varepsilon+\varepsilon^2} &\quad \varepsilon>1
   \end{array}\right. \ .
   \end{equation}
We stress, however, that this estimation  is weaker than the one
obtained from the trace norm of realigned matrix (see Fig.~1).
 \begin{figure}[t]
       \centerline{\includegraphics[width=0.85\linewidth]{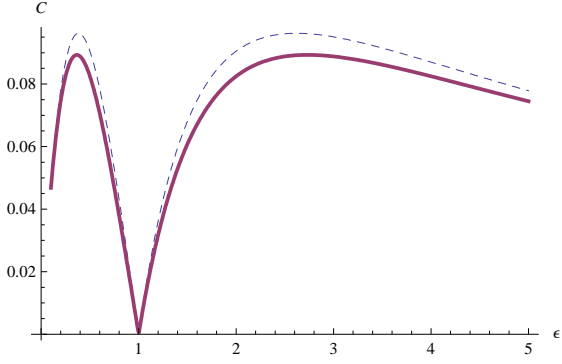}}
 \caption{Two estimations of concurrence as a function of $\varepsilon$. The dashed line is for the estimation based on $||{\cal R}(\rho(\varepsilon))||_1$ due to (\ref{C-thm1}). The solid line is for the estimation based on (\ref{exmp2}).}
 \end{figure}
  }
\end{example}

\begin{example}
{\em Sixia and Yu \cite{SixiaYu}  constructed a family of
entanglement witnesses $W(a)$ for the
   Horodecki states in $\mathbb{C}^3\otimes\mathbb{C}^3$ \cite{HorA}
   ($0<a<1$):
   $$\rho(a)=\frac{1}{8a+1}\left(\begin{array}{ccc|ccc|ccc}
   a&\cdot&\cdot&\cdot&a&\cdot&\cdot&\cdot&a\\
   \cdot&a&\cdot&\cdot&\cdot&\cdot&\cdot&\cdot&\cdot\\
   \cdot&\cdot&a&\cdot&\cdot&\cdot&\cdot&\cdot&\cdot\\
   \hline
   \cdot&\cdot&\cdot&a&\cdot&\cdot&\cdot&\cdot&\cdot\\
   a&\cdot&\cdot&\cdot&a&\cdot&\cdot&\cdot&a\\
   \cdot&\cdot&\cdot&\cdot&\cdot&a&\cdot&\cdot&\cdot\\
   \hline
   \cdot&\cdot&\cdot&\cdot&\cdot&\cdot&\frac12(1+a)&\cdot&\frac12\sqrt{1-a^2}\\
   \cdot&\cdot&\cdot&\cdot&\cdot&\cdot&\cdot&a&\cdot\\
   a&\cdot&\cdot&\cdot&a&\cdot&\frac12\sqrt{1-a^2}&\cdot&\frac12(1+a)
   \end{array}\right) \ .
   $$
   The witness $W(a)$ which detects entanglement of $\rho(a)$ has the following form
   \[
   W(a)\;=\;\1-f(a)V(a)
   \]
   where
   \[
   f(a)\;=\;2\sqrt{(a+2)[(1+8a)^2+a^2(1-a)]}
   \]
   and the real symmetric matrix $V(a)$ reads
   \begin{equation}\label{mV}
   V(a)=\left(\begin{array}{ccccccccc}
   v_{11}&\cdot&v_{13}&\cdot&v_{15}&\cdot&v_{17}&\cdot&v_{19}\\
   \cdot&v_{22}&\cdot&\cdot&\cdot&\cdot&\cdot&v_{28}&\cdot\\
   v_{13}&\cdot&v_{33}&\cdot&\cdot&\cdot&v_{37}&\cdot&v_{39}\\
   \cdot&\cdot&\cdot&v_{44}&\cdot&v_{46}&\cdot&\cdot&\cdot\\
   v_{15}&\cdot&\cdot&\cdot&v_{55}&\cdot&\cdot&\cdot&v_{59}\\
   \cdot&\cdot&\cdot&v_{46}&\cdot&v_{66}&\cdot&\cdot&\cdot\\
   v_{17}&\cdot&v_{37}&\cdot&\cdot&\cdot&v_{77}&\cdot&v_{79}\\
   \cdot&v_{28}&\cdot&\cdot&\cdot&\cdot&\cdot&v_{88}&\cdot\\
   v_{19}&\cdot&v_{39}&\cdot&v_{59}&\cdot&v_{79}&\cdot&v_{99}
   \end{array}\right)
   \end{equation}
For the full list of entries $v_{ij}$ see the Appendix. One has
   \begin{eqnarray*}
  \Tr[W(a)\rho(a)]&=& 1-f(a)\,\Tr(V(a)\rho(a)) \\
  &=&1-f(a)\,\frac{2(2+33a+145a^2+63a^3)}{1+8a}\,.
   \end{eqnarray*}
   Numerical calculations show again that $A_{kl}^{(W(a))}\geq-1$ for $k\neq l$.
Hence, one obtains the following estimation for concurrence
    \begin{equation}\label{exmp3}
    C(\rho(a))\;\geq\;-\Tr[W(a)\rho(a)]/\sqrt{3} \ .
    \end{equation}
Again this estimation is weaker than the one obtained from the
realignment (see Fig.~2).
    \begin{figure}[t]
   \centerline{\includegraphics[width=0.85\linewidth]{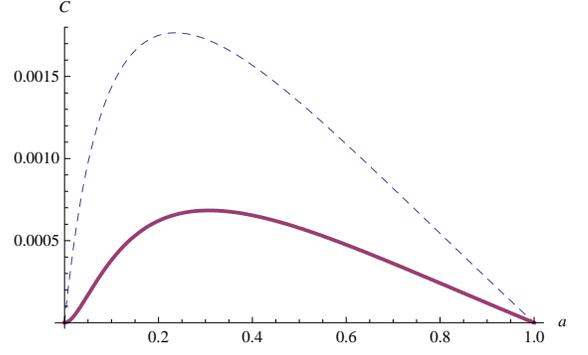}}
   \caption{Two estimations of concurrence as a function of $a$. The dashed line is for the estimation based on $||{\cal R}(\rho(a))||_1$ due to (\ref{C-thm1}). The solid line is for the estimation based on (\ref{exmp3}).}
   \end{figure}
}
\end{example}

\begin{example} {\em  Using the Tang map \cite{Tang} one can construct the following family of
entanglement witnesses
\def\d{\cdot}
\[
W(u)\;=\;\left[\begin{array}{cccc|cccc}
1-u^2/6 &\d&\d&\d&\d&-1&\d&\d\\
\d &1&\d&\d&\d&\d&-2&\d\\
\d&\d&2&\d&u&\d&\d&-2\\
\d&\d&\d&1&\d&\d&\d&\d\\
\hline
\d&\d&u&\d&u^2&\d&\d&-u\\
-1&\d&\d&\d&\d&2&\d&\d\\
\d&-2&\d&\d&\d&\d&2&\d\\
\d&\d&-2&\d&-u&\d&\d&1
\end{array}
\right] \ .
\]
This family detects  Horodecki states $\rho(b)$ $(0<b<1)$ in
$\mathbb{C}^2\otimes\mathbb{C}^4$ \cite{HorA}. Now,
\[ \Tr[W(u)\rho(b)]\;=\;\frac{3-3b-6u\sqrt{1-b^2}+3u^2+2bu^2}{6+42b}
\]
and $W(u)$ detects $\rho(b)$ if and only if $u_1\leq u\leq u_2$, where
\begin{eqnarray*}
u_1 &=& \frac{3\sqrt{1-b^2}-\sqrt{3b(1-b)}}{3+2b}\\
u_2 &=& \left\{\begin{array}{cl}
1 &\;\dST b<\frac{12}{37} \\[2ex]
\dST\frac{3\sqrt{1-b^2}+\sqrt{3b(1-b)}}{3+2b} &\;\dST b\geq\frac{12}{37}\,.
\end{array}\right.
\end{eqnarray*}
Numerical results show that $A_{kl}^{(W(u))}\geq -2$ for $k\neq l$
and hence we define rescaled witness by
\[ \widetilde{W}(u)=
\frac{1}{2}W(u)\ . \] The estimation for concurrence
$$ C(\rho(b))\geq-\Tr[\rho(b)\widetilde{W}(u)]/\sqrt{3} $$ is shown in Fig.~3.
\begin{figure}[t]
\centerline{\includegraphics[width=0.85\linewidth]{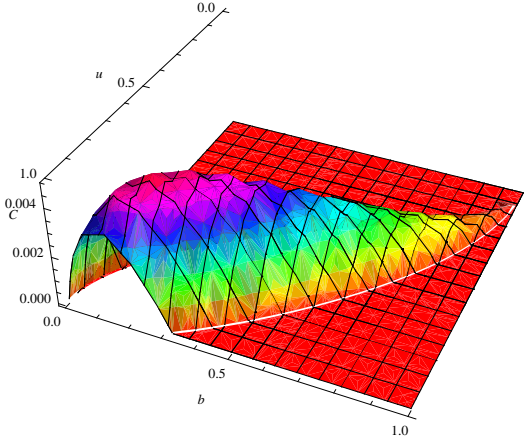}}
\caption{Estimation of concurrence as a function of $b$ and $u$.}
\end{figure}
}
\end{example}

\begin{example} {\em Consider a family of states
in $\mathbb{C}^d\otimes\mathbb{C}^d$ defined by  \cite{Ha,Chr08}
\begin{equation}\label{}
\rho_\gamma= \frac 1 N_\gamma \sum_{i,j=1}^d |i\>\< j| \otimes
A_{ij}^\gamma\,,
\end{equation}
where
\begin{eqnarray*}
A_{11}^\gamma&=&|1\>\<1|+a_\gamma |2\>\<2| +\sum_{\ell=3}^{d-1}|\ell\>\<\ell|+b_\gamma |d\>\<d|\\
A_{ij}^\gamma &=& |i\>\<j|\,,\qquad i\neq j\,,\\
A_{jj}^\gamma &=& S^{j-1}A^\gamma_{11}S^{\dagger j-1}\,,
\end{eqnarray*}
with
\[
a_\gamma=\frac{1}{d}(\gamma^2+d-1)\,,\qquad b_\gamma=\frac{1}{d}(\gamma^{-2}+d-1)\,.
\]
and the normalization factor reads
\[ N_\gamma = d^2-2+\gamma^2+\gamma^{-2} \ .
\]
The operator $S : \mathbb{C}^d \rightarrow \mathbb{C}^d$ is defined
by $S|k\>=|k+1\>$ (mod$\,d$). Note that for $d=3$ the state
$\rho_\gamma$ has a very similar structure to the states
$\rho(\varepsilon)$ (\ref{rhoE}) considered in Example 2. Now, as
was shown in \cite{Ha}, the states $\rho_\gamma$ are detected by a
family of entanglement witnesses
\[
W_{d,k}=\sum_{i,j=1}^d |i\>\<j| \otimes X_{ij}^{d,k}
\]
generalizing those described by (\ref{choi1}) and (\ref{choi2}) (which correspond to $d=3$ and $k=1$).
The $d\times d$ matrices $X_{ii}^{d,k}=(d-k-1)|i\>\<i| +\sum_{\ell=1}^k|i+\ell\>\<i+\ell|$ and $X_{ij}^{d,k}=
-|i\>\<j|$ for $i\neq j$
(all additions  mod$\,d$). Numerical calculations show that
$A_{mn}^{(W_{d,k})}\geq -C_{d,k}$, where for coefficients $C_{d,k}$
we conjecture the following analytic formulae
\[ C_{d,k}=\left\{\begin{array}{cl}
\dST\frac{d-k}{2} &\mbox{for}\; d-2\geq 2k\,,\\[1ex]
\dST\frac{d-k}{2} - \frac 14 &\mbox{for}\; d-2<2k\,,\\[1ex]
1 &\mbox{for}\;k=d-2\,.
\end{array}\right.
\]
Now, for a rescaled witness $\widetilde{W}_{d,k}=W_{d,k}/C_{d,k}$ we
obtain
\[
\Tr(\widetilde{W}_{d,k}\rho_\gamma)=\frac{\gamma^2-1}{d^2-2+\gamma^2+\gamma^{-2}}\cdot\frac{1}{C_{d,k}}
\]
and hence the estimation for concurrence of $\rho_\gamma$ reads
\begin{equation}\label{C-rho-gamma}
C(\rho_\gamma)\geq \sqrt{\frac{2}{d(d-1)}}\cdot\frac{1-\gamma^2}{d^2-2+\gamma^2+\gamma^{-2}}\cdot\frac{1}{C_{d,k}}\,.
\end{equation}
In Fig.~\ref{Ha1} we have shown the estimation of concurrence (\ref{C-rho-gamma}) of $\rho_\gamma$ for different
values of $d$ and maximal $k=d-2$. The case corresponds therefore to the detection of entanglement by the Choi witness in $d=3$ and its natural generalization in $d=4,5$. It is shown that the estimation of concurrence becomes weaker when the dimension $d$ increases.
\begin{figure}[t]
\centerline{\includegraphics[width=0.85\linewidth]{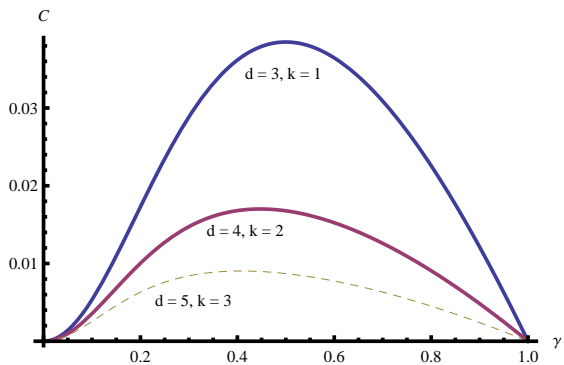}}
\caption{\label{Ha1} Estimation of concurrence (\ref{C-rho-gamma}) as a function of $\gamma$ for different values of $d$ and $k=d-2$, i.e., for the case corresponding to the Choi map for $d=3$.}
\end{figure}

The influence of the parameter $k$ on the estimation of concurrence
for $d=5$ is shown in Fig.~\ref{Ha2}. One can see that the best
estimation gives the witness corresponding to the maximal available
value of $k=d-2$ -- the one which generalizes the Choi witness.
\begin{figure}[t]
\centerline{\includegraphics[width=0.85\linewidth]{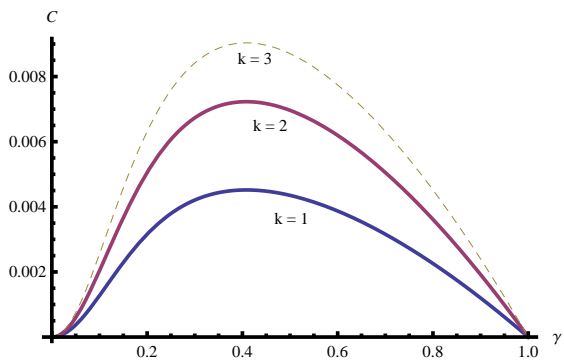}}
\caption{\label{Ha2} Estimation of concurrence (\ref{C-rho-gamma}) as a function of $\gamma$ for $d=5$ and different values of $k$.}
\end{figure}
}
\end{example}

\section{Conclusions}

We shown that each entanglement witness detecting given bipartite
entangled state in $\mathcal{H}_A \otimes \mathcal{H}_B$ provides an
estimation of its concurrence. We analyzed an estimation for
concurrence provided by an (properly rescaled) entanglement witness
for different families of states in various dimensions  and compared
the corresponding estimation of concurrence with other estimations
provided by the trace norm of partial transposition and realignment.
It is shown that typically entanglement witnesses give weaker
estimations than those obtained by realignment but formulae for
estimations are analytic.

We introduced   a  quantity $\lambda$ (cf. formula (\ref{lambda}))
which does provide new characterization of an arbitrary entanglement
witness. This quantity  defines an {\it optimal} rescaling which
gives rise to the best estimation of concurrence.

We analyzed a family
of EWs $W_{d,k}$ in $\mathbb{C}^d \otimes \mathbb{C}^d$. It is shown
that the best estimation is provided by EW corresponding to $k=d-2$.
It turns out that it generalizes the witness based on the Choi map
in $d=3$.

\acknowledgments This work was partially supported by the Polish
Ministry of Science and Higher Education Grant No
3004/B/H03/2007/33 and Grant UMK 370-F.

\section*{Appendix}

The list of entries of the matrix $V(a)$:
\begin{eqnarray*}
v_{11}=v_{66} &=& -(1+a)(1+8a)\\
v_{13}=v_{46} &=& -(1+7a)\sqrt{1-a^2} \\
v_{15}=v_{59} &=& 2(2+a)(1+8a)\\
v_{17}=v_{39} &=& -(1+9a)\sqrt{1-a^2}\\
v_{19} &=& 3(1+a)(1+8a)\\
v_{22}=v_{55} &=& 2+19a+15a^2\\
v_{28} &=& (2+15a)\sqrt{1-a^2}
\end{eqnarray*}
\begin{eqnarray*}
v_{33}=v_{44} &=& (3+a)(1+8a)\\
v_{37} &=& -(1-a)(1+8a) \\
v_{77}=v_{99} &=& 2+17a+17a^2\\
v_{79} &=& (2+17a)\sqrt{1-a^2} \\
v_{88} &=& -2a(1+8a)
\end{eqnarray*}


\begin{thebibliography}{10}
\bibitem{QIT} M. A. Nielsen and I. L. Chuang, {\it Quantum computation
and quantum information}, Cambridge University Press, Cambridge,
2000.

\bibitem{Hor09} R. Horodecki, P. Horodecki, M. Horodecki, K. Horodecki, {\it Quantum entanglement}, Rev. Mod. Phys.
{\bf 81}, 865--942 (2009).
\bibitem{Toth} O. G\"uhne, G. Toth, {\it Entanglement detection}, Phys. Reports {\bf 474}, 1--75 (2009).
\bibitem{Guh07} O. G\"uhne, M. Reimpell, R.F. Werner, {\it Estimating entanglement measures in experiments},
Phys. Rev. Lett. {\bf 98}, 110502 (2007).
\bibitem{Eis} J. Eisert, F.\,G.\,S.\,L. Brandao, K.\,M.\,R. Audenaert, {\it Quantitative entanglement witnesses},
New J. Phys. {\bf 9}, 46 (2007).
\bibitem{Ved98} V. Vedral, M. B. Plenio, M. A. Rippin, and P. L. Knight, {\it Quantifying Entanglement},
Phys. Rev. Lett. {\bf 78}, 2275 (1998).
\bibitem{Ben96} C. H. Bennett, D. P. DiVincenzo, J. A. Smolin, W. K.
Wootters, {\it Mixed-state entanglement and quantum error correction}, Phys. Rev. A {\bf 54}, 3824 (1996).
\bibitem{Woo98} W. K. Wootters, {\it Entanglement of Formation of an Arbitrary State of Two Qubits}, Phys. Rev. Lett. {\bf 80}, 2245 (1998).
\bibitem{Vid} G. Vidal and R. Tarrach, {\it
Robustness of entanglement},  Phys. Rev. A  {\bf 59} 141 (1999).
\bibitem{Steiner} M. Steiner, {\it Generalized robustness of entanglement}, Phys. Rev. A {\bf 67}, 054305 (2003).
\bibitem{Bar01} H. Barnum and N. Linden, {\it Monotones and invariants for multi-particle quantum states}, J. Phys. A: Math. Gen. {\bf 34},
6787 (2001).
\bibitem{Wei03} T.-C. Wei and P. M. Goldbart, {\it Geometric measure of entanglement and applications to bipartite and multipartite quantum states}, Phys. Rev. A {\bf 68}, 042307
(2003).
\bibitem{isotropic} P. Rungta, C.M. Caves, {\it Concurrence-based entanglement measures for isotropic states},
Phys. Rev. A {\bf 67}, 012307 (2003).
 \bibitem{KaiRMP} Kai Chen, S. Albeverio, Shao-Ming Fei, {\it Concurrence-based entanglement measure for Werner States},
  Rep. Math. Phys. {\bf 58}, 325-334 (2006).
  \bibitem{Ter} B. M. Terhal and K. G. H. Vollbrecht, {\it Entanglement of Formation for Isotropic States}, Phys. Rev. Lett. {\bf 85}, 2625 (2000).
\bibitem{Loh} R. Lohmayer, A. Osterloh, J. Siewert, and A. Uhlmann, {\it Entangled Three-Qubit States without Concurrence and Three-Tangle}, Phys. Rev. Lett. {\bf 97}, 260502 (2006).
\bibitem{Datta} A. Datta, S. T. Flammia, A. Shaji, C. M. Caves, {\it Constrained bounds on measures of entanglement},
Phys. Rev. A {\bf 75}, 062117 (2007).
\bibitem{Guh08} O. G\"uhne, M. Reimpell, R.F. Werner, {\it Lower bounds on entanglement measures from incomplete information}, Phys. Rev. A {\bf 77}, 052317 (2008).
\bibitem{AugLew09} R. Augusiak, M. Lewenstein, {\it Towards measurable bounds on entanglement measures},
 Quantum Information Processing {\bf 8}, 493--521 (2009).
 \bibitem{Ma} Zhihao Ma, Fu-Lin Zhang, Dong-Ling Deng, Jing-Ling Chen, {\it Bounds of concurrence and their relation
 with fidelity and frontier states}, Phys. Lett. A {\bf 373}, 1616--1620 (2009).
 \bibitem{deVicente} J.I. de Vicente, {\it Lower bounds on concurrence and separability conditions}, Phys. Rev. A.
 {\bf 75}, 052320 (2007).
 \bibitem{Min04} F. Mintert, M. Ku\'s, A. Buchleitner, {\it Concurrence of mixed bipartite quantum states in
 arbitrary dimensions}, Phys. Rev. Lett. {\bf 92}, 167902, (2004).
 \bibitem{Min07} F. Mintert, {\it Entanglement measures as physical obserwables}, App. Opt. B {\bf 89}, 493-497 (2007).
 \bibitem{Min07a} F. Mintert, {\it Concurrence via entanglement witnesses}, Phys. Rev. A {\bf 75}, 052302 (2007).
  \bibitem{Min07b} F. Mintert, A. Buchleitner, {\it Observable entanglement measure for mixed quantum states},
 Phys. Rev. Lett. {\bf 98}, 140505 (2007).
 \bibitem{Chr} D. Chru\'sci\'nski, A. Kossakowski, {\it On the structure of entanglement witnesses and new class of positive indecomposable maps}, Open Sys. Information Dyn. {\bf 14}, 275--294 (2007).
 \bibitem{Ou} Yong-Cheng Ou, Heng Fan, and Shao-Ming Fei,
{\it Proper monogamy inequality for arbitrary pure quantum states},  Phys. Rev. A {\bf 78}, 012311 (2008).
   \bibitem{Chen} Kai Chen, S. Albeverio, Shao-Ming Fei, {\it Concurrence of Arbitrary Dimensional Bipartite Quantum States}, Phys. Rev. Lett. \textbf{95}, 040504 (2005).
   \bibitem{R1} Kai Chen, Ling-An Wu, {\it A matrix realignment method for recognizing entanglement},
   Quantum Information and Computation {\bf 3}, No.~3 (2003) 193--202.
   \bibitem{R2} Cheng-Jie Zhang, Yong-Sheng Zhang, Shun Zhang, Guang-Can Guo,
   {\it Entanglement detection beyond the cross-norm or realignment criterion}, Phys. Rev. A {\bf 77},
   060301(R) (2008).
   \bibitem{Zhang} Cheng-Jie Zhang, Yong-Sheng Zhang, Shun Zhang, Guang-Can Guo,
   {\it Optimal entanglement witnesses based on local orthogonal observables}, Phys. Rev. A {\bf 76},
   012334 (2007).
   \bibitem{Brandao} F. G. S. L. Brandao, {\it Quantifying entanglement with witness operators},
   Phys. Rev. A {\bf 72}, 022310 (2005).
   \bibitem{Breuer} H.-P. Breuer, {\it Separability criteria and bounds for entanglement measures}, J. Phys. A: Math. Gen. \textbf{39}, 11847 (2006).
   \bibitem{TerHor} B.M. Terhal, P. Horodecki, {\it Schmidt number for density matrices}, Phys. Rev. A {\bf 61}, 040301 (2000).
   \bibitem{Jur09} J. Jurkowski, D. Chru\'sci\'nski, A. Rutkowski, {\it A class of bound entangled states of two qutrits}, Open Sys. Information Dyn. \textbf{16}, 235 (2009).
   \bibitem{SixiaYu} Sixia Yu and Nai-le Liu, {\it Entanglement Detection by Local Orthogonal Observables}, Phys. Rev. Lett. \textbf{95}, 150504 (2005)
   \bibitem{HorA} P. Horodecki, {\it Separability criterion and inseparable mixed states with positive partial transposition}, Phys. Lett. A \textbf{232}, 333 (1997).
   \bibitem{Tang} Wai-Shing Tang, {\it On positive linear maps between matrix algebras}, Lin. Alg. Appl. \textbf{79}, 33 (1986).
   \bibitem{Ha}  K.-C. Ha, {\it Atomic positive linear maps in matrix algebras}, RIMS (Kyoto) 34, 591 (1998).
   \bibitem{Chr08} D. Chru\'sci\'nski and A. Kossakowski, J. Phys. A: Math. Theor. {\bf 41}, 145301 (2008).




     \end{thebibliography}
\end{document}